\documentclass[prl,checkin,preprintnumbers,amsmath,amssymb]{revtex4}

\newcommand{\mbf}[1]{\mbox{\boldmath $#1$}}

\def\0{\mbox{\mbf 0}}

\def\BibTeX{{\rm B\kern-.05em{\sc i\kern-.025em b}\kern-.08em
    T\kern-.1667em\lower.7ex\hbox{E}\kern-.125emX}}

\usepackage{graphicx}

\usepackage{dcolumn}

\usepackage{bm}

\begin{document}

\preprint{APS/123-QED}

\title{Synchronization of Extended Systems From Internal Coherence}

\author{Gregory S. Duane}
\affiliation{
National Center for Atmospheric Research, P.O. Box 3000, Boulder, CO
80307}
%

%

\begin{abstract}

A condition for the synchronizability of a pair of PDE systems, coupled through a finite
set of variables, is commonly the existence of internal synchronization or internal coherence 
in each system separately.   The condition was previously illustrated in a forced-dissipative system, and is here extended to 
Hamiltonian systems, using an example from particle physics. Full synchronization is precluded by Liouville's
theorem. 
A form of synchronization weaker than ``measure synchronization" is manifest as the positional
coincidence of coherent oscillations (``breathers" or ``oscillons") in a pair of coupled scalar field models in an
expanding universe with a nonlinear potential, and does not occur with a variant of the model that does not
exhibit oscillons.

\end{abstract}



\maketitle


%

%

The phenomenon of synchronized chaos, initially explored in low-order
ODE systems\cite{PC,Afr,FY}, has been extended to PDEs that describe a variety of systems
of physical interest\cite{KocPRL97}. Chaos synchronization extends the 
paradigm of synchronization of regular oscillators that is ubiquitous in Nature \cite{Strog}.
One seeks an
understanding of the internal properties of a chaotic physical system
that will allow a
pair of such systems, loosely coupled, to synchronize, despite sensitive dependence on
initial conditions. Spatially extended systems offer richer possibilities for 
relationships that fall short of full synchronization than do ODE systems.
In geophysical example previously studied
\cite{DT04}, it was seen that slaving of small scales, a relationship that defines
an {\it inertial manifold}, was crucial to synchronizability.  Here, it is
suggested that more general inter-scale relationships, as may give rise to
 coherent structures within each system
separately, are required for  the synchronizability of the pair.
The connection is illustrated in a particle physics context - a toy model of
a scalar field in the expanding early universe \cite{Farhi}, a Hamiltonian system
without an attractor. Weak scale relationships allow oscillons (breathers)
to persist, and give rise to a new form of synchronization defined by the coincidence
in position of oscillons in two coupled scalar field models. Conversely,
where the dynamics do not admit such coherent structures, vestiges of
synchronization are lost.

In the non-Hamiltonian meteorological example 
described in \cite{DT04}, which we first review, two geophysical
fluid systems, representing planetary-scale wind patterns, are coupled only
through their medium-scale Fourier components.  Each system is given 
by a potential vorticity equation
\begin{equation}
\label{eqpv0}
\frac{Dq_i}{Dt}\equiv
\frac{\partial q_i}{\partial t}+J(\psi_i,q_i)=F_i+D_i
\end{equation}
where the streamfunction $\psi$ is the fundamental dynamical variable, the Jacobian
$J(\psi,\cdot){=}\frac {\partial \psi}{\partial x}\frac{\partial \cdot}
{\partial y}{-}\frac {\partial \psi}{\partial y}\frac{\partial \cdot}{\partial x}$ gives
the
advective contribution to the Lagrangian derivative $D/Dt$, there are two horizontal layers $i{=}1,2$, and the potential
vorticity $q$, which generalizes angular
momentum, is a derived variable defined in terms of $\psi$ in \cite{DT04}.
Potential vorticity is conserved on a moving
parcel, except for forcing $F_i$ and dissipation $D_i$.

Two models of the form (\ref{eqpv0}), ${D q^A}/{D t}{=}F^A{+}D^A$ and
${D q^B}/{D t}{=}F^B{+}D^B$
were coupled diffusively through one of the forcing terms:
\renewcommand{\k}{{\vec k}}
\begin{eqnarray}
\label{eqFcoup}
F^A_\k&=&\mu^c_\k[q^B_\k-q^A_\k] + \mu^{\rm ext}_\k  [q^*_\k - q^A_\k] \nonumber \\
F^B_\k&=&\mu^c_\k[q^A_\k-q^B_\k] + \mu^{\rm ext}_\k  [q^*_\k - q^B_\k] 
\end{eqnarray}
where the flow has been decomposed spectrally and the subscript $\k$ on each
quantity indicates the wave number $\k$ spectral component (suppressing the index
$i$). A background flow $q^*$ forces each system separately. The set of coefficients $\mu^c_\k$ was 
chosen to couple the two channels only in some medium range
of wavenumbers.
Band-limited coupling defined by $\mu^c_\k$ replaces
the coupling of two PDE systems at a discrete set of points as in \cite{KocPRL97}.

The two systems synchronize over time in Fig. \ref{figQGsync},
where the contours of $\psi$ are streamlines that define the flow.
The synchronization is manifest as the coincidence of structures of
meteorological significance (``blocking patterns" that interrupt the flow) in both space and time. That 
coincidence is robust
against significant differences in the two systems \cite{DT04}.

The large scales need not be coupled because of well known dynamical relations between
scales (``inverse cascade") in 2D turbulence \cite{DTPRL}. But the
stronger result \cite{DT04} that synchronization occurs without coupling of the smallest 
scales is explained simply: The smallest-scale components are thought not to be independent dynamical variables
but are  functions
of the medium and large-scale components, defining a dynamically invariant
{\it inertial manifold}.  Where the set of variables that are coupled is smaller than a minimal independent set,
synchronization does not occur.
Partial synchronization of forced-dissipative systems will
usually be easy to arrange, as previously found by Kocarev et al. \cite{KocPRL97}, 
since {\it approximate inertial manifolds} (AIM's) exist for almost all
parabolic PDEs \cite{refAIMs}.


In Hamiltonian systems, the subject of the current work, inertial manifolds cannot
exist, since the stability of such a manifold would imply a collapse of phase space volumes for trajectories that start off
the manifold, contradicting
Liouville's Theorem.  
A simple example is the Klein-Gordon equation in an expanding background geometry, 
in one space and one time 
dimension, with cyclic boundary conditions, with a nonlinear potential of a type that gives rise to oscillons,
possibly representing the first coherent structures in the universe.  The field satisfies
\begin{equation}
\label{eqKG}
\frac{\partial^2 \phi}{\partial t^2}+H\frac{\partial \phi}{\partial
t}-e^{-2Ht} \frac{\partial^2 \phi}{\partial x^2}+V'(\phi)=0
\end{equation}
where $H$ is a Hubble constant, the potential
$V$ is given by $V(\phi)= (1/2) \phi^2 - (1/4) \phi^4 + (1/6) \phi^6$
for units in which the scalar field mass $m=1$,
and the prime denotes the derivative with respect to $\phi$. 
The equation (\ref{eqKG}) is derived by replacing derivatives in the standard Klein-Gordon equation
by covariant derivatives for a Robertson-Walker metric describing expansion
with Hubble constant $H$ \cite{Weinberg}, giving a term in $\partial\phi/\partial t$ which formally resembles friction but which
in fact preserves the Hamiltonian structure. Indeed, the system (\ref{eqKG}) is
derivable from a
time-dependent Hamiltonian density 
\begin{equation}
\label{Ham}
{\cal H} = (1/2) e^{-Ht} [(\phi_x)^2 + \pi^2] + e^{Ht} V(\phi)
\end{equation}
where $\pi \equiv e^{Ht}\dot\phi$ is the canonical momentum that is conjugate to $\phi$.
Liouville's theorem applies even with the time dependence, so that
neither an inertial manifold, in the strict
sense, nor an AIM with a usefully small approximation bound can exist. 

Oscillating coherent structures appear as regions of high energy density at final time, in the numerical
integration shown in Fig. \ref{figosc}a,  
after initializing with thermal noise as in \cite{Farhi}. 


Consider two  systems (\ref{eqKG}) coupled diffusively,
through some Fourier components of the field only, according to
\begin{equation}
\label{eq2KG}
\frac{\partial^2 \phi^{A,B}}{\partial t^2}+H\frac{\partial \phi^{A,B}}{\partial
t}-e^{-2Ht} \frac{\partial^2 \phi^{A,B}}{\partial x^2}+V'(\phi^{A,B})-F^{A,B}=0 
\end{equation}
\begin{equation}
\label{eqforcosc}
F_k^A= c_k \left(\frac{\partial \phi^B_k}{\partial t} - \frac{\partial \phi^A_k}{\partial t}\right)\hspace{.5in}
F_k^B=c_k \left(\frac{\partial \phi^A_k}{\partial t} - \frac{\partial \phi^B_k}{\partial t}\right)
\end{equation}
with the coupling coefficients $c_k=0$ vanishing above a threshold value $|k|> k_0$ and set to a large value for $|k|\le k_0$, 
so that corresponding large scale components in the
two systems are effectively clamped.  
For coupling of the lowest wavenumber modes, up to wavelengths of about twice the final
oscillon width,
oscillons in the two systems were 
found to occur at mostly the same locations, though their amplitudes differed, 
as shown in Fig. \ref{figoscsync}a.  
In contrast, if the two systems were left completely uncoupled, but shared initial conditions
over a range of scales, and only the smallest scales, well below the oscillon
widths, were initialized differently, then oscillons formed at locations that were
uncorrelated (Fig. \ref{figoscsync}b).  Apparently, there is a ``butterfly effect", as in meteorology, through which the
small scales have a large impact on the positions of formation of coherent structures.\cite{fn1}  
But the dynamical evolution of these structures then partially slaves a portion of the
small-scale sector (as with shock waves) and proceeds independently of the remaining portion.


With severely attenuated initial noise (Fig. \ref{figoscinitatten}a), oscillon synchronization
occurs with an even narrower range of coupled wavenumbers. The large-scale components
(top 32 modes) of the same oscillons are shown in the bottom portion of Fig.
\ref{figoscinitatten}b. For four of the five oscillons shown, the
oscillon positions are coincident with the local maxima of the truncated
field. For the remaining one (the leftmost), the oscillon is displaced from
the local maximum of the truncated field, but is still coincident with the
corresponding oscillon in the other system (Fig.\ref{figoscinitatten}a).
The small-scale Fourier components, in both situations, are slaved to the
large-scale components insofar as they determine oscillon position.


Coincidence of oscillon positions suggests {\it measure synchronization},
the weak form of synchronized chaos in which the trajectories of two coupled
systems become the same when the systems are coupled, without a requirement that
the states of the systems are the same at a given instant of time\cite{meassync,Afrnote}.  
Measure synchronization
is characteristic of jointly Hamiltonian coupled systems. Here, the coupled
oscillon systems do not quite attain measure synchronization, since the corresponding
oscillons differ in amplitude, and the configuration (\ref{eq2KG}) which can be written:
\begin{eqnarray}
\label{nonHam}
\dot\phi_k^{A,B}&=&e^{-Ht}\pi_k =\partial{\cal H}^{A,B}/\partial \pi_k^{A,B}\\
\dot\pi_k^{A,B}&=&-\partial{\cal H}^{A,B}/\partial \phi_k^{A,B} + c_k(\pi_k^{B,A}-\pi_k^{A,B}) \label{nonHamb}
\end{eqnarray}
is not derivable from
a joint Hamiltonian. (It would be Hamiltonian if the second term in (\ref{nonHamb})
were to be replaced by $c_k(\phi_k^{B,A} - \phi_k^{A,B})$.)
But even without the joint Hamiltonian structure, complete synchronization is not likely achievable by
coupling a finite number of modes, if the remaining infinite number are 
not slaved.  

In a comparable system without oscillons, there are no vestiges of synchronization.
Correlations between corresponding modes in the coupled systems are displayed
in Table \ref{tabcor}a for the oscillon system, and in Table \ref{tabcor}b for 
an alternate pair of systems with potential
$V=V_{\mbox {alt}}(\phi)= (1/2) \phi^2 + (1/4) \phi^4 + (1/6) \phi^6$
in (\ref{eqKG}), for which no oscillons form, as seen in Fig.
\ref{figosc}b. 
(Oscillons occur in one case and not the
other because the $-\phi^4$ term in the first case
gives a flatter potential, so that larger amplitude oscillations have lower
frequencies and decouple from the faster, smaller travelling wave solutions that would
otherwise cause them to dissipate.) While there are small but sigificant correlations
between some of the uncoupled modes in Table \ref{tabcor}a, corresponding to the
small partially-slaved portion of the small-scale sector, 
no significant correlations appear between corresponding small-scale
components that are not coupled  in Table \ref{tabcor}b.   The correlations in the
case with oscillons are expected to extend to much shorter wavelengths
in longer simulations that approach the oscillon lifetimes, as the oscillons decrease in width (in
comoving coordinates) and background fluctuations decrease in amplitude \cite{Farhi}.



The synchronization results for oscillons suggest a unifying principle governing
synchronization in both Hamiltonian and forced-dissipative systems: 
Two complex systems of either type can be made to synchronize, with a 
restricted
set of coupled variables, if and only if each system exhibits synchronization internally.  
Sufficiency
of the condition follows from transitivity if the internal synchronization is
exact: Any 
external coupling
that does not destroy the internal relationships and causes some pairs of variables in the two
systems to agree will also cause all variables that internally synchronize with
the coupled variables to agree with their counterparts in the other system.  
In the cases considered here, the condition also appears necessary.  
Localized coherent structures like oscillons, as well as long-range connections such as the Atlantic-Pacific
``teleconnection" described previously \cite{DTPRL,DT04}, can be regarded as approximate forms
of internal synchronization. 
Where a master set of variables, in Fourier space, synchronizes with the entire remaining infinite set,
an inertial manifold exists.
Local coherence has
been described as synchronization of a more limited type in PDEs \cite{Hale} 
and in lattices of coupled maps \cite{cohsync}. 


The internal synchronization condition also provides guidance as to the choice of coupled variables.
If the extended systems are ``truth" and a ``computer model" to which truth is
coupled in one direction only, via observations, as in meteorology \cite{geosumm}, then a set of variables 
should be observed that is synchronized with a maximal set of other
variables internally.

An internal coherence criterion applies to ODE systems, consistently with Pecora and Carroll's criterion of
negative ``conditional Lyapunov exponents" in the Lorenz system \cite{PC}. 
Lorenz X can be said to approximately synchronize with Lorenz Y, along the near-planar Lorenz attractor, in agreement 
with the well known fact that
sufficiently strong coupling of either X or Y in one system to the corresponding variable of a second Lorenz system
will synchronize the two systems. Coupling of the Z variables, that do not correlate internally with either X or Y,  
will not do so.
But the internal coherence criterion is much more
useful for PDEs that describe spatially extended systems, both because such systems are less tractable 
analytically - a full set of conditional Lyapunov exponents is hard to compute - and because the coherent
structures are more meaningful physically. 


For Hamiltonian systems coherent structures seem to bear strongly on synchronizability and on
the form of synchronization.
The ubiquity of coherent structures in solutions to nonlinear PDE's is the basis on which potentially broad relevance
is claimed for the current work. Solitons, for instance, in a pair of PDE systems that exhibit them,
might also be made to coincide in position, and to move in synchrony, if only a restricted set of
corresponding Fourier components of the two fields are coupled. The question is whether the internal dynamics
that allow the structures to exist in each system would cause the needed parts of the uncoupled components to follow. The
phenomenon, if it exists, might be used for secure communications in the same manner as the systems considered
in \cite{KocPRL97}.
 
The present generalization differs from measure synchronization \cite{meassync}
in several notable respects: First, in an ergodic system with trajectories that define a uniform measure,
such as the system with modified potential $V_{\mbox {alt}}$ that exhibits no oscillon behavior, measure synchronization
is trivial. Second, as already pointed out, positional coincidence of coherent structures that
differ in amplitude or detailed shape is more general than measure synchronization.  Third, there is no
requirement that the {\it combined} system be Hamiltonian to exhibit the weak form of synchronization
described here. The impossibility of strict synchronization follows not from the joint Hamiltonian
structure, but from the impossibility of slaving or approximately slaving all uncoupled variables
in each system separately.


The new types of synchronization appear to be equivalent, in their details, to detailed inter-scale relationships, within
each system internally, that in the present case allow oscillons to persist and to stably maintain their positions.
The partial agreement of small scales between the two synchronized systems, sufficient to force coincidence in oscillon 
position, when large scales are
clamped, indeed defines a partial slaving that counters
the butterfly effect.  For consistency
with Liouville's Theorem, there must be a compensating expansion of the remaining part of the
uncoupled modes in phase space, as entropy is cast off to scales that are yet smaller. 
Details of the partial slaving 
remain to be worked out. 
The existence and the form of synchronization may 
provide diagnostics for the slaving relationships.


{\it Acknowledgements:} This work was supported in part by NSF
Grant 0327929. The author thanks Joe Tribbia, Roger Temam, Noah Graham,
and Jeff Weiss for useful discussions, and especially thanks Alan Guth
for also conducting some of the numerical experiments.


\newpage

%
%

\begin{table}
 \caption{Correlations between coefficients of corresponding Fourier components of the independent dynamical 
variables in the $A$ and $B$ subsystems of the coupled scalar field system (\ref{eq2KG}) (a) and for the same
system with the potential $V_{\mbox {alt}}$ (b). The Fourier coefficients, indexed by $n$, are partitioned into coupled
modes with wavenumber $k=[(n+1)/2]$, $0\le n<128$ (in cycles/domain-length), and several ranges of uncoupled modes.  Error bars for
the first range of uncoupled modes are at $2\sigma$, where $\sigma$ was computed as the standard error of the
mean based on a further partitioning into odd and even $n$. Uncoupled modes correlate significantly only with
a nonlinear potential that supports oscillons.}
 \label{tabcor}
 \begin{tabular}{rlcccc}
 \multicolumn{2}{c}{}&\multicolumn{2}{c} {a) correlations with}  
          & \multicolumn{2}{c} {b) correlations with}   \\
 \multicolumn{2}{c}{} & \multicolumn{2}{c}{$V=\phi^2/2 - \phi^4/4 +
 \phi^6/6$}
            & \multicolumn{2}{c}{\protect\hspace{.25in}$V_{\mbox{alt}}=  \phi^2/2 + \phi^4/4 + \phi^6/6$}   \\   
 \multicolumn{2}{c}{} & \multicolumn{1}{c}{$\phi$}
           & \multicolumn{1}{c}{$\partial\phi/\partial t$}  
           & \multicolumn{1}{c}{$\phi$} & \multicolumn{1}{c}{$\partial\phi/\partial t$}  \\   
       \hline
 coupled modes: & $0\le n < 128$  & 1.00  & 1.00  & 1.00 & 1.00    \\
 uncoupled modes:& $128 \le n < 256$ & 0.31 $\pm$ 0.04  & 0.18 $\pm$ 0.18 & -0.06
		                           $\pm$ 0.04 &  0.17 $\pm$ 0.18    \\
 uncoupled modes:& $256 \le n < 384$ & 0.08  & 0.00  &  0.06  & -0.14                      \\
 uncoupled modes:& $384 \le n < 512$ & 0.02 &  0.04  &  0.04  &  0.00              \\
all uncoupled modes: & $128 \le n < 16258$  & 0.08 & 0.00 & 0.00 & 0.01                              \\
  \\ \hline\\
  \end{tabular}
 \end{table}

\protect\newpage

\begin{figure}
\caption{Streamfunction $\psi$ (in units of $1.48\times 10^9 m^2s^{-1}$, averaged over layers
$i=1,2$) describing
the  flow at initial (a,b) and final (c,d) times, in a 
parallel
channel model with coupling of medium scale modes 
for which
$|k_x| > k_{x0}=3$ or $|k_y| > k_{y0}=2$, and $|k|\le 15$, for the
indicated numbers $n$ of time steps in a numerical integration. Parameters
are as in \cite{DT04}.
 Synchronization
occurs by the last time shown (c,d), despite differing initial conditions.
The ``blocking patterns" in the boxed areas coincide \protect\cite{DTPRL}.}
\label{figQGsync}
\end{figure}

\begin{figure}
\caption{Energy density 
$\rho = (1/2) e^{-Ht}(\phi_x)^2 + (1/2) e^{Ht}(\phi_t)^2 
+ e^{Ht} V(\phi)$ vs. position $x$ for a numerical simulation of
(\ref{eqKG}), suggesting localized oscillons (a), and a simulation
of the same equation, but with a different potential
$V(\phi)= (1/2) \phi^2 + (1/4) \phi^4
+ (1/6) \phi^6$, for which oscillons do not occur, shown for 
comparison (b).}
\label{figosc}
\end{figure}

\begin{figure}
\caption{a)The local energy density $\rho$ vs. x for two simulations of the oscillon system (\ref{eqKG}),
coupled according to (\ref{eq2KG}) and (\ref{eqforcosc}), with coupling coefficient $c_k=2$  
for $k \le 64$ and $c_k=0$ otherwise, at final time.  
($\rho$ for the second system (dashed line) is also shown reflected across the 
x-axis for ease in comparison.) The coincidence of oscillon positions is apparent.;
b) Local energy density $\rho$ vs. x at final time for two simulations of the oscillon system, plotted
as in a), but with common initialization of all modes with  $k < 0.8\times2^{14}$, and no subsequent coupling between the two systems. 
Oscillon positions
appear uncorrelated.}
\label{figoscsync}
\end{figure}

\begin{figure}
\caption{a)The local energy density $\rho$ vs. x for two simulations of the oscillon system (\ref{eqKG}),
coupled according to (\ref{eq2KG}) and (\ref{eqforcosc}), with coupling coefficient $c_k=2$  
for $k \le 32$ and $c_k=0$ otherwise, at final time, displayed as in Fig.
\ref{figoscsync}. The initial noise level was severely attenuated as compared
to the thermal initialization used in the simulation in Fig. \ref{figoscsync}:
the amplitude of the $n^{\mbox{th}}$ Fourier component at initial time
was multiplied by $(1/n)^{0.35}$. 
b) Local energy density $\rho$ vs. x at final time for one of the two
simulations shown in panel a), with the part of $\rho$ corresponding to
$k \le 32$ shown reflected in the negative-$y$ portion of the panel.}
\label{figoscinitatten}
\end{figure}

\newpage
\setcounter{figure}{0}

\begin{figure}
\vspace{3in}
 \begin{minipage}{7in}
 \begin{minipage}{7in}
 \hspace{.7in} channel A \hspace{1.8in} channel B 
 \end{minipage}
 \begin{minipage}{7in}
  \bigskip \hspace{0.2in} \raisebox{.5ex}{n=0} 
 \end{minipage}
 \begin{minipage}{7in}
 a) \resizebox{.3\textwidth}{!}{\includegraphics{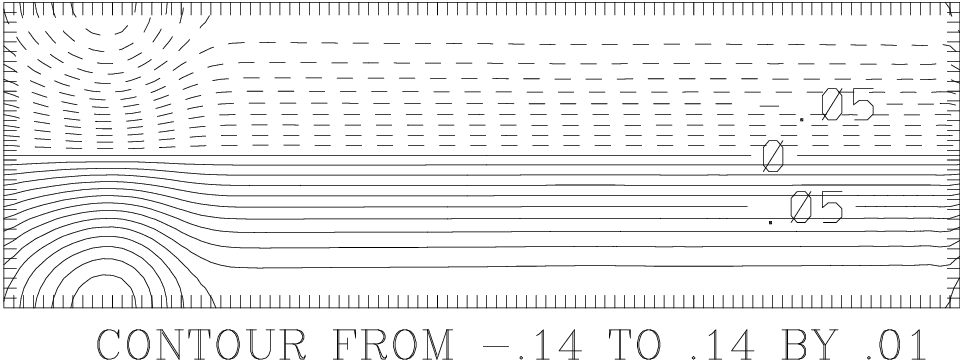}} \hspace{.2in}
   b) \resizebox{.3\textwidth}{!}{\includegraphics{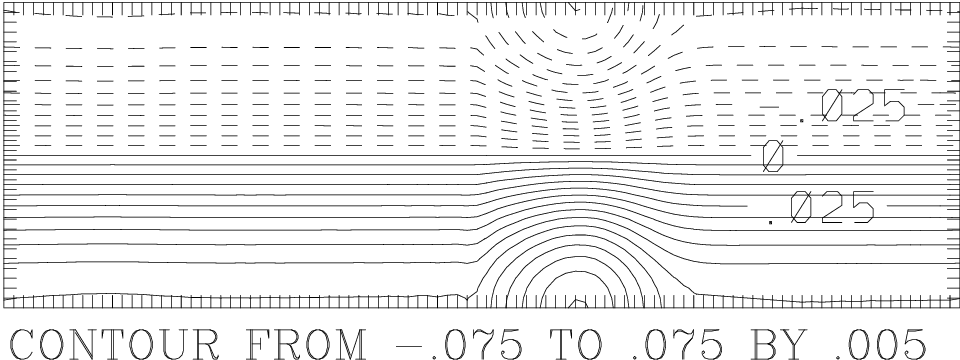}}
 \end{minipage}
 \begin{minipage}{7in}
  \bigskip \hspace{0.2in} \raisebox{.5ex}{n=2000} 
 \end{minipage}
 \begin{minipage}{7in}
 c) \resizebox{.3\textwidth}{!}{\includegraphics{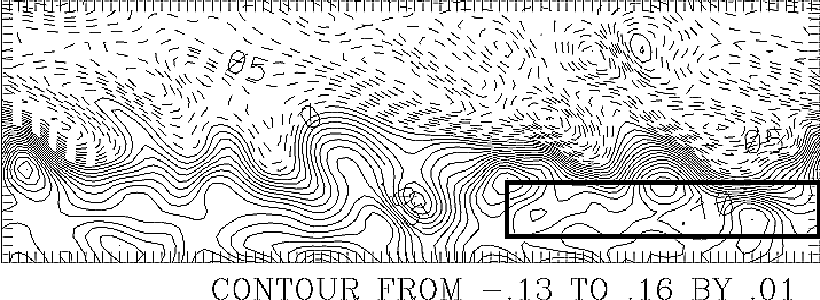}} \hspace{.2in}
   d) \resizebox{.3\textwidth}{!}{\includegraphics{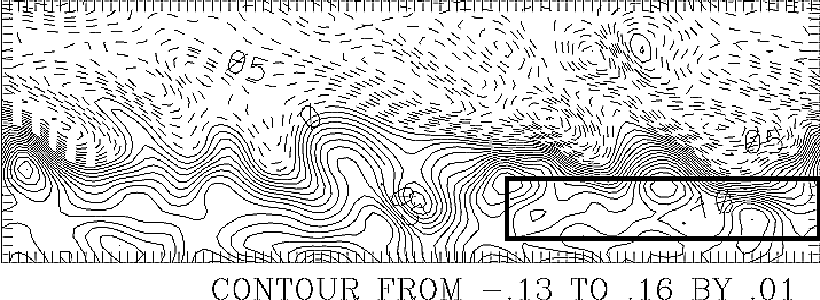}}
 \end{minipage}
  \begin{minipage}{7in}
   \end{minipage}
  \end{minipage}   
\caption{}
\end{figure}

\newpage

\begin{figure}
\vspace{3in}
a)   \resizebox{.9\textwidth}{!}{\includegraphics{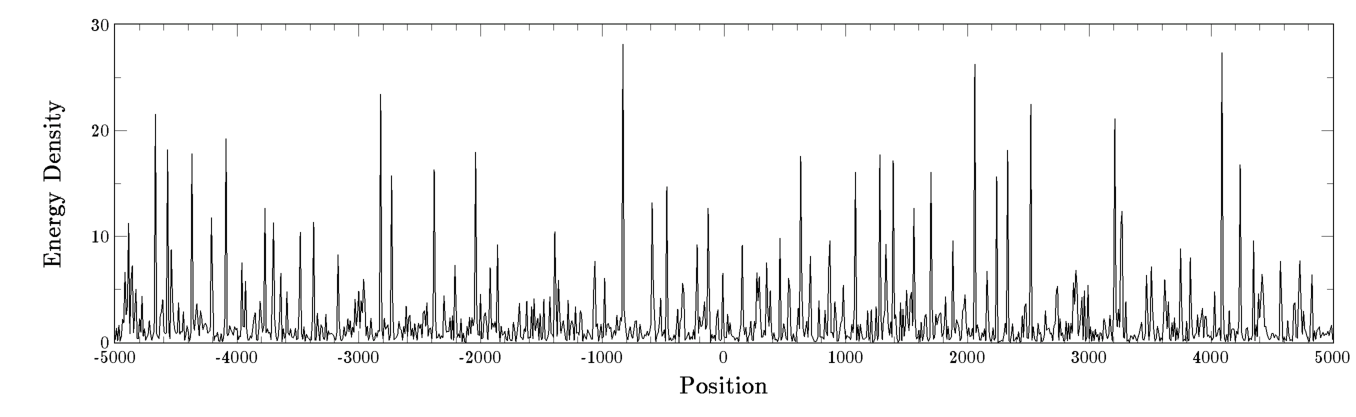}}\\ 
b)   \resizebox{.9\textwidth}{!}{\includegraphics{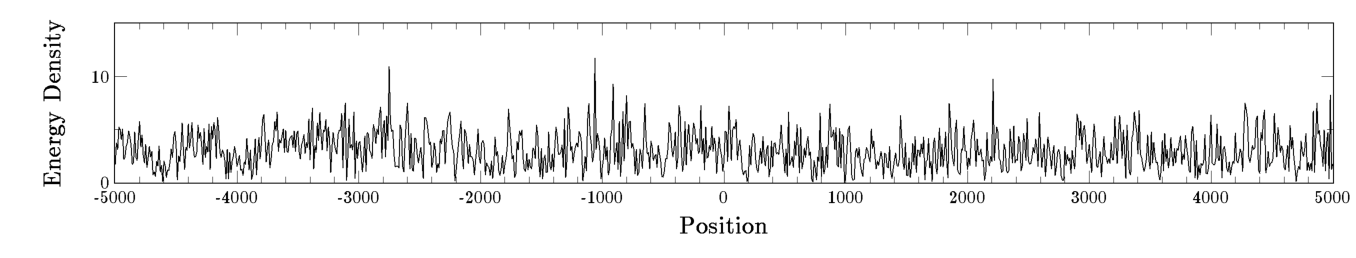}}
\caption{}
\end{figure}

\newpage

\begin{figure}
\vspace{3in}
a)\resizebox{.4\textwidth}{!}{\includegraphics{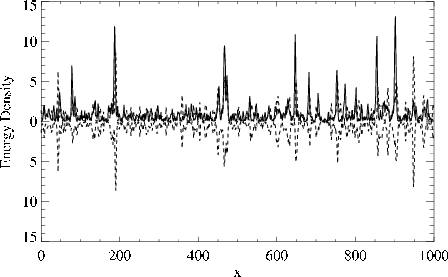}}\hspace{.5in}
b)\resizebox{.4\textwidth}{!}{\includegraphics{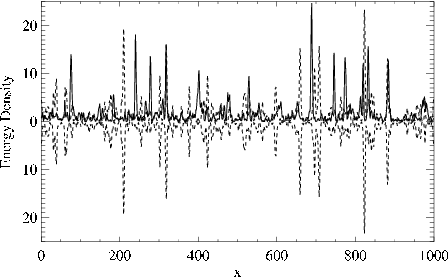}}
\caption{}
\end{figure}

\newpage

\begin{figure}
\vspace{3in}
a)\resizebox{.4\textwidth}{!}{\includegraphics{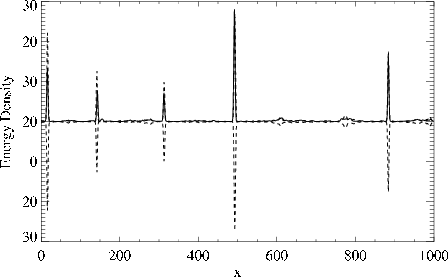}}\hspace{.5in}\\
\vspace{.5in}
b)\resizebox{.4\textwidth}{!}{\includegraphics{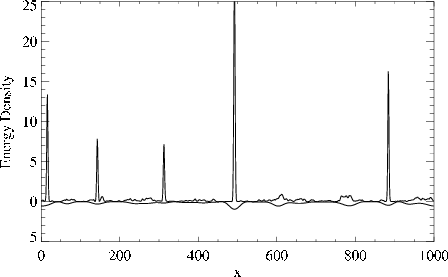}}
\caption{}
\end{figure}


\begin{thebibliography}{99}

\bibitem{PC} L.M. Pecora and T.L. Carroll, Phys. Rev. Lett. {\bf 64}, 821 (1990).

\bibitem{Afr} V.S. Afraimovich, N.N. Verichev, and M.I. Rabinovich, Inv. VUZ
Radiofiz. RPQAEC {\bf 29}, 795 (1986).

\bibitem{FY} H. Fujisaka and T. Yamada, Prog. Theor. Phys. {\bf 69}, 32
(1983).

\bibitem{KocPRL97} L. Kocarev, Z. Tasev, and U. Parlitz, Phys. Rev. Lett. {\bf 79}, 51 (1997).

\bibitem{Strog} S. Strogatz, {\it Sync: The Emerging Science of Spontaneous Order},
Theia, New York (2003).

\bibitem{DT04}
G.S. Duane and J.J. Tribbia, J. Atmos. Sci. {\bf 61}, 2149 (2004).

\bibitem{Farhi} E. Farhi, N. Graham, A.H. Guth, N. Iqbal, R.R. Rosales, and
N. Stamatopoulos, Phys. Rev. D {\bf 77}, 085019 (2008).

\bibitem{PC97} L.M. Pecora, T.L. Carroll, G.A. Johnson,
and D.J. Mar, Chaos {\bf 7}, 520 (1997).

\bibitem{DTPRL}
G.S. Duane and J.J. Tribbia, Phys. Rev. Lett. {\bf 86}, 4298 (2001).

\bibitem{refAIMs} R. Temam, {\it Infinite Dimensional Dynamical
Systems in Mechanics and Physics}, Springer-Verlag, New-York,
Applied Mathematical Sciences Series {\bf 68}, (1988).

\bibitem{Weinberg} S. Weinberg, {\it Gravitation and Cosmology: Principles
and Applications of the General Theory of Relativity}, Wiley, New York (1972).

\bibitem{fn1} The butterfly effect here differs essentially from bubble nucleation 
as in Coleman, S., Phys. Rev. D {\bf 15}, 2929 (1977). 
The oscillons do not grow from
a seed, but shrink, in the coordinate system of (\ref{eqKG}). Initial and
final oscillon widths are both much larger than the ``seed" scale. 

\bibitem{meassync} A. Hampton and D.H. Zanette, Phys. Rev. Lett. {\bf 83}, 2179 (1999).

\bibitem{Afrnote} Measure synchronization was first reported as ``non-isochronic"
synchronization in Ref. \protect\cite{Afr}.


\bibitem{Hale} J.K. Hale, J. Dynamics and Diff. Eq. {\bf 9}, 1 (1997).

\bibitem{cohsync} K. Kaneko (Ed.), {\it Theory and Applications of Coupled Map 
Lattices}, Wiley, New York (1993).

\bibitem{geosumm} G.S. Duane, J.J. Tribbia, and J.B. Weiss, Nonlin. Processes
in Geophys. {\bf 13}, 601 (2006); G.S. Duane and J.J. Tribbia, 
Ch. 17 in {\it Nonlinear Dynamics in the
Geosciences}, ed. A. Tsonis, Springer (2007).

\end{thebibliography}
\end{document}